\documentstyle[12pt,epsf]{article}								 
\begin{document} 
\vspace*{3cm}
\begin{center}
{\large\bf 
$T$-odd correlation in the  $K_{l 3 \gamma }$-decay}
\end{center}

\vspace*{1cm}
\begin{center}
V.V. Braguta$^{\dagger}$, A.A. Likhoded$^{\dagger\dagger}$, A.E. Chalov$^{\dagger}$
\end{center}

\vspace*{1cm}
\begin{center}
$^{\dagger}$ {\it Moscow Institute of Physics and Technology,  Dolgoprudny, 141700
Russia} \\
$^{\dagger\dagger}$ {\it Institute for High Energy Physics,  Protvino, 
142284 Russia}\\ {\em e-mail: andre@mx.ihep.su}
\end{center}

\vspace*{2cm}
\underline{\bf Abstract}
\vspace*{0.5cm}
\newline
The dependence  of the $K^+ \to \pi^0 l^+ \nu_l \gamma$ decay width 
on the  $T$-odd kinematical variable,  
$ \xi = { \vec q \cdot [\vec p_l \times \vec p_\pi]} / {M^3}$,
is studied at the tree and one-loop levels of Standard Model.
It is shown that at the tree level this decay width is the even
function of  $\xi$, while the odd contribution arises due to the 
electromagnetic final state interaction. This contribution
is determined by imaginary parts of one-loop diagrams. 
The calculations performed show that the $\xi$-odd contribution
to the $K^+ \to \pi^0 e^+ \nu_e \gamma$ and
$K^+ \to \pi^0 \mu^+ \nu_\mu \gamma$ decay widths 
is four orders of magnitude smaller than even contribution coming 
from the tree level of SM.

\newpage
\section*{Introduction}

\noindent
The study of  rare radiative $K$-meson decays provide an interesting 
possibility to search for effects of a new physics beyond the Standard
Model (SM). In particular, the search for 
new $CP$-violating interactions is of special interest.
Contrary to SM, where the $CP$-violation is caused by the presence of the complex phase 
in the  CKM matrix,  the $CP$-violation in extended models
can naturally arise due to the presence of, for instance, new charged Higgs bosons,
which have  complex couplings  to fermions [1], hypothetical tensor interactions [2], 
etc. $CP$-violating effects can be probed with experimental observables, which are
especially  sensitive to  $T$-odd contributions. Such observables are
the  rate dependence on $T$-odd correlation ($\xi =\frac{1}{M_K^3}
\vec p_{\gamma}\cdot [\vec p_{\pi} \times \vec p_l ]$)  in the 
$K^{\pm}\to\pi^0 \mu^{\pm}\nu\gamma$ process [3] and transverse muon polarization
in the $K^{\pm}\to \mu^{\pm}\nu\gamma$ decays [4].    
 The experiments conducted thus far do not provide the sensitivity level, which is 
 necessary to analyze the differential distributions in the 
 $K^{\pm}\to\pi^0 \mu^{\pm}(e^{\pm })\nu\gamma$ decays. However, 
new perspectives are connected with the planned OKA experiment [5], where
expected statistics of $\sim 7.0\cdot 10^5$ events for the 
$K^+\to\pi^0 \mu^+\nu\gamma$ decay allows one to perform detailed
analysis of the data and either probe new effects or put strict bounds 
on the parameters of  extended models. 

Searching for possible $T$-violating effects caused by
new interactions in the  $K^+\to\pi^0 \mu^+\nu\gamma$ decays
it is especially important to estimate the SM contribution to
$\xi$-distribution, which is induced by electromagnetic final state
interaction and which is natural background for new 
interaction contributions.

The Weinberg model with three Higgs doublets [1,6] is especially interesting
for the search of possible $T$-violation. This model allows one to have 
complex Yukawa couplings that leads to extremely interesting phenomenology. 
It was shown [3] that the study of the $T$-odd correlation
in the $K^+\to\pi^0 \mu^+\nu\gamma$ process allows one either to probe the terms, which 
are linear in $CP$-violating couplings, or strictly confine  the Weinberg model
parameters. 

In this paper, in the framework of SM we analyze the 
$K^+ \to \pi^0 l^+ \nu_l \gamma$ decay width dependence on the kinematical
variable $\xi = { \vec q \cdot [\vec p_l \times \vec p_\pi]} / {M^3}$. 
In general case, the width differential distribution, 
$\rho ( \xi ) = d \Gamma /d \xi$, can be represented as the sum
of the even, $f_{even}$, and odd, $f_{odd}$, functions of $\xi$. 
At the tree level of SM the odd part, $f_{odd}$, does not contribute
to the width distribution. We will show later that 
this effect is a direct consequence of the following fact: in the chiral 
perturbation theory the formfactors contributing to the matrix element  
do not have imaginary parts. However, the SM radiative corrections
due to the electromagnetic final state interaction 
lead to the appearance of formfactors imaginary parts [7], that, in its turn,
results in a nonvanishing $\xi$-odd contribution in the 
 $K^+ \to \pi^0 l^+ \nu_l \gamma$ decay width  distribution.
In this paper we analyze this effect at the one-loop level of SM.
The matrix element of the $K^+ \to \pi^0 l^+ \nu_l \gamma$ decay
is calculated in the leading approximation of the 
chiral perturbation theory, i.e. up to the terms of $O( p^4 )$ [8].

   To probe the $T$-odd effect we introduce, besides  $f_{odd}$, 
the $\xi$-asymmetrical physical observable, which is defined as follows
\begin{equation}
A_\xi =\frac {N_+ - N_-} {N_+ + N_-}\;,
\end{equation}
where $N_+$ and $N_-$ are the numbers of events with $\xi >0$ and $\xi <0$,
correspondingly. One can see that while the $A_\xi$ nominator depends on
$f_{odd}( \xi )$ only the  denominator is proportional to $f_{even}( \xi )$, that
makes this variable sensitive to  $\xi$-odd effects.

 As we will show later,  the ``background'' SM one-loop contribution to  
 $f_{odd}$ is severely suppressed with respect to  $f_{even}$ 
 ($f_{odd}/f_{even} \sim 10^{-4}$). This allows us to state that
 proposed observables, $A_\xi$ and $f_{odd}$, sensitive to $T$-odd contributions,
 provide good chance to search for the $CP$-violating effects beyond SM.

Another variable, sensitive to the $CP$-violation,
is the transverse muon polarization, $P_T$, which can be observed in the
$K^+ \to \pi^0 \mu^+ \nu$ and $K^+ \to \mu^+ \nu \gamma$ decays [4,7,9]. 
As in the case of $\xi$-dependence of the $K^+ \to \pi^0 l^+ \nu_l \gamma$ rate, 
the presence of nonvanishing transverse polarization in SM
is caused by the electromagnetic final state interaction. Though 
the $P_T$ value is sensitive to $T$-odd effects, its measurement in the experiment
seems to be cumbersome [10]. 
As for the $A_\xi$ and $f_{odd}$ variables,
their experimental measurement is much more easier, that is one of the main advantages  
of these variables in comparison with the transverse muon polarization.
The low event rate of the processes, where these values can be measured,
is considered to  be as the disadvantage of these variables. However,
the anticipated statistics on the $K^+ \to \pi^0 \mu^+ \nu\gamma $
process in the OKA experiment definitely 
allows one to use these observables to search 
for new  $CP$-violating contributions.

In the first section we analyze the 
$K^+ \to \pi^0 l^+ \nu_l \gamma$ decay width dependence on 
the  $T$-odd correlation at the tree level of SM. In Section~II
we calculate SM contributions to the  $T$-odd correlation, induced by
one-loop diagrams. Last section contains the discussion and
conclusions.

\section{$'$-odd correlation at the tree level of SM}

\noindent  
The Feynman diagrams contributing to the 
$K^+(p) \to \pi^0(p') l^+(p_l) \nu_l(p_\nu) \gamma(q)$ decay at the tree level of SM
are shown in Fig~1. 
The tree-level amplitude for this process can written as [8]:
\begin{eqnarray}
T= \frac {G_F} {\sqrt {2}} e V^*_{us} {\epsilon ^{\mu} (q)}^* \biggr( 
(V_{\mu \nu}-A_{\mu \nu}) \overline u (p_{\nu}) \gamma^{\nu} (1- \gamma_5 ) v(p_l)+ \nonumber
\\ + \frac {F_{\nu}} {2 p_l q} \overline u (p_{\nu}) \gamma^{\nu} (1- \gamma_5 )
(m_l- \hat p_l - \hat q) \gamma_{\mu} v (p_l)
\biggl) ,
\end{eqnarray}
where
\begin{eqnarray}
V_{\mu \nu}=i \int d^4 x e^{i q x} \langle  \pi^0 (p ')|
T V_{\mu}^{em} (x) V_{\nu}^{4-i5}(0)  
| K^+ (p) \rangle \\
A_{\mu \nu}=i \int d^4 x e^{i q x} \langle  \pi^0 (p ')|
T V_{\mu}^{em} (x) A_{\nu}^{4-i5}(0)  
| K^+ (p) \rangle, 
\end{eqnarray}
and  $ F_{\nu}$ is the matrix element of the $K^+_{l3}$-decay:
\begin{eqnarray}
F_{\nu}=\langle  \pi^0 (p ')|
 V_{\nu}^{4-i5}(0)  
| K^+ (p) \rangle \;.
\end{eqnarray}
Here $p'$, $p_l$,  $q$, $p_\nu$ ¨ $p$ are the pion, lepton,
$\gamma$-quantum, neutrino, and kaon four-momenta, correspondingly. 
In the leading approximation of the chiral
perturbation theory $ A_{\mu \nu}=0 $ and expressions for
$V_{\mu \nu }$ and $ F_{\mu}$ can be written as
\begin{eqnarray}
F_{\mu}&=&\frac 1 {\sqrt 2}  (p+p')_\mu \nonumber \\
V_{\mu \nu}&=&V_1 (g_{\mu \nu}-\frac {W_{\mu} q_{\nu}} {qW}) + 
V_2 (p'_{\mu} q_{\nu}-\frac {p'q} {qW} W_{\mu} q_{\nu}  )+
\frac {p_{\mu}} {pq} F_{\nu} \nonumber \\
W_{\mu}&=&(p_l+p_{\nu})_{\mu}	\nonumber \\
V_1&=&\frac 1 {\sqrt 2} ,~~
V_2=-\frac 1 {\sqrt 2 p q} \;. \nonumber
\end{eqnarray}
So, the matrix element of the decay can be rewritten in the following form:
\begin{eqnarray}
T=\frac{G_F}{2}e V^*_{us} \epsilon^{\mu}(q)^* \cdot\overline{u}(p_\nu)(1+\gamma_5)\cdot
\hspace*{7.cm}\nonumber\\
\left((\hat{p}+\hat{p}')\Biggl(\frac {p_\mu}{(pq)}-\frac{(p_l)_\mu}{(p_lq)}\Biggr)
-(\hat{p}+\hat{p}')\frac{\hat{q}\gamma_\mu}{2(p_lq)}+\Biggl(\gamma_\mu-\frac{\hat{q}p_\mu}{(pq)}
\Biggr)\right)u(p_l)\;,
\end{eqnarray}
and the $K^+ \to \pi^0 l^+ \nu \gamma$ decay partial width 
can be calculated by integrating over the phase space.

In Fig.~2 we present the differential distribution
of the decay partial width in the $K$-meson rest frame over 
the three-momenta of final particles and the angle between the lepton
and $\gamma$-quantum directions, calculated at the tree level of SM. 
For the case of the electron channel (see Fig.~2a)
the bulk of the width value is collected 
in the region of small values of the lepton
and $\gamma$-quantum momenta, maximal values of the pion momenta, and
small angles between lepton and $\gamma$-quantum momenta.

In the case of the muon channel (see Fig.~2b)
the bulk of the width is collected 
in the region of intermediate values of lepton momentum,
small values of the  $\gamma$-quantum momentum, maximal values of the pion
momentum, and small angles between the lepton and
$\gamma$-quantum directions. 

Imposing the kinematical cuts on the $\gamma$-quantum energy and
lepton-$\gamma$-quantum scattering angle in the kaon rest frame,
$E_\gamma >30 $~MeV and $\theta_{\gamma \l}>20^0$, which are typical
for the current and planned kaon experiments, one gets the following 
branching values:
\begin{eqnarray}
\hbox{Br}(K^+ \to \pi e^+ \nu_e \gamma)&=&3.18 \cdot 10^{-4}\;, 	\nonumber   \\ 
\hbox{Br}(K^+ \to \pi \mu^+ \nu_\mu \gamma)&=&2.15 \cdot 10^{-5}\;, \nonumber 
\end{eqnarray}
which are in a good agreement with earlier calculations (see, for instance, [8]) and
existing experimental results [11]. 

Looking for possible $CP$-odd contributions we will investigate
the decay width distribution over variable
$\xi={ \vec q \cdot [\vec p_l \times \vec p_\pi]} / {M^3}$, which 
changes the sign under  $CP$- or $T$-conjugation,
\begin{equation}
\rho ( \xi ) = \frac {d \Gamma} {d \xi}\;,
\end{equation}
This distribution is an ``indicator'' for the  $T$-violation effects.
The  $\rho(\xi)$ function can be rewritten as 
\begin{displaymath}
\rho=f_{even}(\xi)+f_{odd}(\xi)\;,
\end{displaymath}
where $f_{even}(\xi)$ and $f_{odd}(\xi)$ are the even and odd
functions of $\xi$, correspondingly.
The function $f_{odd}(\xi)$ can be represented as follows
\begin{equation}
f_{odd}=g(\xi^2) \xi\;.
\end{equation}
It is evident, that after integration of the $\rho ( \xi)$ function
over whole region  of $\xi$ only the $f_{even}( \xi)$ function contributes
to the total width. In Fig.~3 we present 
the $\rho ( \xi)/\Gamma_{total}$ distributions for the  
$K^+ \to \pi^0 \mu^+ \nu_l \gamma$ and $ K^+ \to \pi^0 e^+ \nu_l \gamma$ decays.
Indeed, one can see from Fig.~3 that at the tree level of SM, where there are
no any $T$-odd contributions, the distributions, as one could expect,
are strictly symmetric with respect to the line $\xi =0$, i.e. the numbers of
events of the  
$K^+ \to \pi^0 l^+ \nu_l \gamma$ decay with $\xi >0$ and $\xi < 0$ are equal.
This fact can be explained by following: in the case of the tree approximation of SM
the matrix element squared is expressed via scalar products of final particles momenta 
only, and, consequently, there are no  contributions linear over $\xi$. 
So,  the $\rho ( \xi)$ function  is essentially even function of $\xi$.

Analysing the $K^+ \to \pi^0 l^+ \nu_l \gamma$ data, it is useful
to introduce, besides the $\rho ( \xi)$ distribution,
the integral asymmetry, which is  defined as
\begin{equation}
A_\xi =\frac {N_+ - N_-} {N_+ + N_-}\;,
\end{equation}
where $N_+$ and $N_-$ are the numbers of decay events with $\xi >0$ and $\xi <0$. It is
easy to see that the  $A_\xi$ nominator depends on  $f_{odd}( \xi )$ only, which
makes this variable highly sensitive to  $T$-odd effects beyond SM.

\section{ $'$-odd correlation in SM due to the final state 
interaction}

\noindent
Nonvanishing value of the  $A_\xi$-asymmetry as well as odd
contribution to $\rho (\xi)$ can arise in SM  due to the 
electromagnetic final state interaction at the level of one-loop diagrams. 
The most general expression for the
$K^+ \to \pi^0 l^+ \nu_l \gamma$ decay amplitude with account for 
the electromagnetic radiative corrections (implying the gauge invariance) 
can be written as follows:
\begin{eqnarray}
T_{one-loop} &= & \frac {G_F} {\sqrt 2} e V_{us}^* \epsilon_\nu^* \bar u (p_\nu)
(1+\gamma_5)\cdot  \nonumber \\ 
&\cdot & \biggl ( C_1 (p^\nu- \frac {p q} {p_l q} p_l^\nu)+
C_3 ((p')^\nu- \frac {p' q} {p_l q} p_l^\nu)+ 
C_5   (p^\nu- \frac {p q} {p_l q} p_l^\nu) \hat {p'} \nonumber \\  
&& + C_7   ((p')^\nu- \frac {p' q} {p_l q} p_l^\nu ) \hat {p'}+ 
C_9  ( \hat q p^\nu-(p q) \gamma^\nu) \nonumber \\
&& + C_{10}  ( \hat q p_l^\nu-(p_l q) \gamma^\nu)+  
C_{11} ( \hat q (p')^\nu-(p' q) \gamma^\nu)+C_{12} \hat q \gamma^\nu \nonumber \\
&& +  C_{13} \hat {p'} ( \hat q p^\nu-(p q) \gamma^\nu ) +C_{14} \hat {p'} ( \hat q p_l^\nu-(p_l q) \gamma^\nu) \nonumber \\ 
&& +  C_{15} \hat {p'} ( \hat q (p')^\nu-(p' q) \gamma^\nu)+
C_{16} \hat {p'} \hat q \gamma^\nu
\biggr ) v (p_l)\; \label{todd}, 
\end{eqnarray}
Where the $C_i$ coefficients are the kinematical factors, which are 
due to one-loop diagram contributions. The matrix element squared
with account for the one-loop contributions can be rewritten in the following form:
\begin{equation}
|T_{one-loop}|^2 = |T_{even}|^2+ |T_{odd}|^2\;,\label{matrix}
\end{equation}
where 
\begin{eqnarray}
|T_{odd}|^2 &=&  -2 G_F^2  e^2 |V_{us}|^2 m_K^4 \xi \biggl ( 
 \mbox{Im}( C_1 ) m_l (2 \frac {1} {p_l q} - 4 \frac { p q} {(p_l q)^2} )  \nonumber \\  
&& - \mbox{Im}( C_3 ) m_l (2 \frac {1} {p_l q}+4 \frac { p' q} {(p_l q)^2} ) \nonumber\\
&& + \mbox{Im} ( C_5 )	(4+2 m_l^2 \frac { p q } {(p_l q)^2} + \frac {1} {p_l q} 
 (2 m_K^2 - 2 m_\pi^2 +4 p p'\nonumber \\
 && -4 p p_l - 4 p q -4 p' p_l- 4 p' q))   \nonumber \\  
 && + \mbox{Im} ( C_7 )	(2 m_l^2 \frac {p' q} {(p_l q)^2}+ 4 \frac { m_\pi^2 } {p_l q})+
 \mbox{Im} ( C_9 )	(8 \frac {p p_l} {p_l q}- 8 \frac {m_K^2} {p q})  \nonumber \\  
 && +\mbox{Im} (C_{10}) ( 8 \frac {m_l^2} {p_l q} + 8 \frac {p_l q} {p q} -
 8 \frac {p p_l} {p q}-8)+  \mbox{Im} (C_{11}) (8 \frac {p' q} {p q}+ 
 8 \frac {p' p_l} {p_l q} -
 8 \frac {p p'} {p q})  \nonumber \\  
 && +\mbox{Im} (C_{12}) ( 4 \frac {m_l} {p q} -
 8 \frac {m_l} {p_l q})+
 \mbox{Im} ( C_{13}) m_l (4 \frac {m_K^2} {p q} - 4 \frac {p p_l} {p_l q})
  \nonumber \\  
 && +\mbox{Im} (C_{14}) m_l (4+4 \frac {p p_l} {p q}-4 \frac {m_l^2} {p_l q}-
 4 \frac {p_l q} {p q}) \nonumber \\
 &&+\mbox{Im}( C_{15} ) m_l (4 \frac {p p'} {p q}-4 \frac {p' q} {p q} - 
 4 \frac {p' p_l} {p_l q} )
  \nonumber \\  
 && +\mbox{Im}(C_{16}) (-8+ 4 \frac {m_K^2} {p q} - 4 \frac {m_\pi^2} {p q} +
 8 \frac {p p'} {p q} \nonumber \\&& -8 \frac {p p_l} {p q}- 8 \frac {p' p_l} {p q}-
 8 \frac {p' q} {p q} + 4 \frac {m_l^2} {p_l q} + 8 \frac {p_l q} {p q})
\biggr ) \label{dmatrix}
\end{eqnarray}
As one can see from Eqs. (\ref{matrix}) and (\ref{dmatrix}), 
the nonvanishing contribution to  $f_{odd}( \xi)$ and 
$A_\xi$ (linear over $\xi$) is determined by the one-loop
electromagnetic corrections, which lead to the appearance 
of imaginary parts of  the $C_i$ formfactors.

To calculate the formfactor imaginary parts one can use the
$S $-matrix unitarity [7]:
\begin{displaymath}
S^+ S=1\;.
\end{displaymath}
Using $ S=1+i M$, one gets
\begin{equation}
M_{f i}-M_{i f}^*=i \sum_n M^*_{n f} M_{n i}, 
\end{equation}
where $i,\: f, \: n$ indices correspond to the initial, final, and intermediate
states of the particle system.
Further, using the  $T$-invariance of the matrix element one has
\begin{eqnarray}
\mbox{\mbox{Im}} M_{f i}&= &\frac 1 2 \sum_n M^*_{n f} M_{n i}  \nonumber \\
M_{f i}&=&(2 \pi)^4 \delta ( P_f-P_i ) T_{f i} \nonumber
\end{eqnarray}

one-loop diagrams, which describe the electromagnetic corrections
to the $K^+ \to \pi l^+ \nu_l \gamma$ process and lead to 
imaginary parts of the formfactors in (\ref{todd}), thus contributing
to $f_{odd}( \xi)$, are shown in Fig.~3.
Using Eq.~(2) one can write down the imaginary parts of
these diagrams, which give the nonvanishing contribution to $f_{odd}( \xi)$.
It is useful to split the whole set of one-loop diagrams
in to two groups. The first group contains the diagrams shown in
Figs.~4a, 4c, 4e. The imaginary part of these diagrams
can be written as follows: 
\begin{eqnarray}
\hbox{Im}T_1=\frac{  \alpha}{2 \pi}\frac{G_F}{\sqrt{2}}eV_{us}^*
\overline u(p_{\nu})(1+\gamma_{5}) \int \frac{d^3 k_{\gamma}}{2 \omega_{\gamma}}
\frac{d^3 k_l}{2 \omega_l}\delta(k_{\gamma}+k_l-q-p_l)\cdot  \nonumber\\
\cdot \hat R_\mu
(\hat k_l-m_l)\gamma^{\mu}\frac{\hat q+ \hat p_{l}-m_{l}}{(q+p_{l})^2-m_{l}^2}
\gamma^{\delta} \varepsilon^*_{\delta} v(p_{l})	
\end{eqnarray}
The second group includes the diagrams shown in Figs.~4b, 4d, 4f. 
The corresponding imaginary part is
\begin{eqnarray}
\hbox{Im}T_2=\frac{  \alpha}{2 \pi}\frac{G_F}{\sqrt{2}}eV_{us}^*
\overline u(p_{\nu})(1+\gamma_{5}) \int \frac{d^3 k_{\gamma}}{2 \omega_{\gamma}}
\frac{d^3 k_l}{2 \omega_l}\delta(k_{\gamma}+k_l-q-p_l) \cdot \nonumber\\
\cdot \hat R_\mu
(\hat k_l-m_l)\gamma^{\delta} \varepsilon^*_{\delta} 
\frac{ \hat k_{\mu}-\hat q-m_{l}}{(k_{\mu}-q)^2-m_{l}^2}
\gamma^{\mu} v(p_{l}), 
\end{eqnarray}
where
\begin{eqnarray}
\hat R_\mu=(V_{\mu \nu}-A_{\mu \nu}) \gamma^\nu - \frac {F_\nu} {2 p_l q}
\gamma^\nu ( \hat p_l+ \hat q - m_l ) \gamma_\mu
\end{eqnarray}
The details of the integrals calculation entering Eqs. (14), (15), 
and their dependence on kinematical parameters are given in Appendix~1.
Expressions for imaginary parts of the $ C_i $ formfactors
are given in Appendix~2.

\section{Results and discussions}

\noindent
Before discussing the numerical results, let us note that
considering one-loop diagrams we neglected their contributions
to the even part of the  $\xi$-distribution, as these contributions
are considerably smaller than nonzero contribution to $f_{even}$ from
the tree approximation of SM. However, in the case of $f_{odd}$ the 
tree SM contribution is equal to zero, thus the contributions to 
$f_{odd}$ coming from one-loop diagrams become essential.
Analysing the  $K^+ \to \pi^0 \l^+ \nu_l \gamma$ width dependence 
on the kinematical variable $\xi $ we separately consider 
two decay channels, $K^+ \to \pi^0 e^+ \nu_e \gamma$ and
$K^+ \to \pi^0 \mu^+ \nu_\mu \gamma$, since the functional
$\xi$-dependence of the width in these two case is essentially different. 

\vspace*{0.5cm}
\underline{\bf $K^+ \to \pi^0 e^+ \nu_e \gamma$}

\vspace*{0.3cm}
\noindent
In Fig. 5a we show the $\xi$-odd contribution
to the differential width distribution, which
is induced by the imaginary parts of one-loop 
diagrams shown in  Fig.~3.
In the kinematical region of the $\xi$ parameter
the value of the width distribution varies in the interval
of $(-2.0 \div 2.0)\cdot 10^{-6}$, and the sign of
$f_{odd}$ is opposite to the sign of $\xi$. As the total
$\xi$-distribution is the sum of the even and odd parts, it leads
to the fact that in experiment one will observe the surplus
of the events with negative $\xi$ values.
The asymmetry value for this channel is
\begin{displaymath}
A_\xi (K^+ \to \pi e^+ \nu_e \gamma) =-0.59 \cdot 10^{-4}\;. 
\end{displaymath}

\vspace*{0.5cm}
\underline{\bf $K^+ \to \pi^0 \mu^+ \nu_\mu \gamma$}

\vspace*{0.3cm}
\noindent
In Fig. 5b we present the $\xi$-odd contribution
to the differential width distribution for the muon decay channel.
The characteristic variation interval for this distribution is
$(-4.0 \div 4.0)\cdot 10^{-7}$, but the sign of $f_{odd}$ coincides
with the sign of $\xi$. 
This results in the surplus of the events with positive $\xi$ values.
The asymmetry value for this channel is
\begin{displaymath}
A_\xi (K^+ \to \pi \mu^+ \nu_\mu \gamma) =1.14 \cdot 10^{-4}\;.
\end{displaymath}

\vspace*{0.5cm}
\noindent
This difference between the $f_{odd}$ behaviour in cases of
electron and muon channels can be explained as follows: 
for the muon decay channel the contributions from 
imaginary parts of the  $C_1$, $C_{12}$, $C_{13}$, and $C_{14}$
formfactors become essential, while in the case of the 
electron channel their contributions are negligible
(these contributions are proportional to mass of the lepton).

It should be noted that the difference 
in the $f_{odd}$ behaviour for the electron and muon channels 
could be used to disentangle the SM radiative and
new physics contributions: in extended models, where the $CP$-violation
can arise at tree level, the sign of  the $\xi$-dependence is insensitive
to the lepton flavor, as it takes place, for instance,
in the Weinberg model [1].

We would like to underline that for the both decay channels
the  $f_{odd}$ value is four orders of magnitude smaller than 
the tree contribution of SM.  It allows to state
that  $\xi$-odd effects are severely suppressed in SM.
Thus, the ``background'' SM contribution to the odd part of the
$\xi$-dependence leaves the  ``window'' to discover new $CP$-violating
effects in these decays up to the level of $10^{-4}$.

Analysing the situation with the integral asymmetry $A_\xi$ one sees that
for reliable observation of  $\xi$-odd effects  from the asymmetry only
one should have the data sample for these decays at least at the level
of  $10^8$ events. In this respect the analysis of differential 
$\xi$-distribution seems to be very important.

\section*{Acknowledgements}

\noindent
The authors would like to thank L.B. Okun for the critical remarks
and interest to this calculations.
The authors would like to acknowledge useful discussions an valuable remarks
by V.A. Rubakov, A.K. Likhoded, and  V.F. Obraztsov. 
This work was supported, in part, by the RFBR, grants
99-02-16558 and 00-15-96645, and RF Ministry of Education, grant …00-3.3-62.

\newpage
\section*{References}

\vspace*{0.5cm}
\noindent
\begin{description}
\item[1.] S. Weinberg,  Phys. Rev. Lett. {\bf 37}, 651 (1976).
\item[2.] R.J. Tesarek,  hep-ex/9903069. 
\item[3.] A. Likhoded, V. Braguta, and A. Chalov,  in press (see also, 
 hep-ex/0011033).
\item[4.] A. Likhoded, V. Braguta, and A. Chalov,  hep-ph/0105111. 
\item[5.] V.F. Obraztsov, Nucl. Phys. B (Proc. Suppl.) {\bf 99}, 257 (2001). 
\item[6.]  J.F. Donoghue and B. Holstein, Phys. Lett. B {\bf 113}, 382 (1982); 
L. Wolfenstein, Phys. Rev. {\bf 29}, 2130 (1984);
G. Barenboim et al.,  Phys. Rev. D {\bf 55}, 24213 (1997);
M. Kobayashi, T.-T. Lin, and Y. Okada,  Prog. Theor. Phys. {\bf 95}, 361 (1996);
S.S. Gershtein et al., Z. Phys. C {\bf 24}, 305 (1984);
R. Garisto and G. Kane,  Phys. Rev. D {\bf 44}, 2038 (1991);
G. Belanger and C.Q. Cheng,  Phys. Rev. D {\bf 44}, 2789 (1991).
\item[7.] L.B. Okun and I.B. Khriplovich,  Sov. Journ, Nucl. Phys. {\bf 6 }, 821 (1967).
\item[8.] J. Bijnens, G. Ecker, and J. Gasser, Nucl. Phys. B {\bf 396},
81 (1993).
\item[9.]  A.R. Zhitnitskii, Sov. J. Nucl. Phys.  {\bf 31}, 529 (1980);
see also, C.Q. Geng and J.N. Ng,  Phys. Rev. D {\bf 42}, 1509 (1990);
C.H. Chen, C.Q. Geng, and C.C. Lih,  hep-ph/9709447;
G. Hiller and G. Isidori,  Phys. Lett. B {\bf 459}, 295 (1999).
\item[10.]  See, for instance, Yu.G. Kudenko,  hep-ex/00103007.
\item[11.]  Review of Particle Physics,  Euro. Phys. Journ. C {\bf 15},
501 (2000).
\end{description}

\newpage
\section*{Appendix 1}

\noindent
Calculating the integrals, which contribute to Eqs. (14) and (15),  we 
use the following notations:
$$
P=p_l+q
$$
$$
d \rho =\frac {d^3 k_{\gamma}} {2 \omega_{\gamma}}
\frac {d^3 k_{l}} {2 \omega_{l}}\delta(k_\gamma+k_{l}-P) 
$$
We present below either the explicit expressions for integrals,
or the set of equations, which being solved, give the parameters,
entering the integrals.
\begin{eqnarray}
J_{11}&=&\int d \rho =\frac  {\pi} 2 \frac {P^2-m_{l}^2} {P^2}\;,\nonumber \\
J_{12}&=&\int d \rho \frac 1 {(p k_{\gamma})}=
\frac {\pi}	{2 I} \ln \biggl( \frac {(P p)+I} {(P p)-I} \biggr)\;,\nonumber 
\end{eqnarray}
where
$$
I^2=(P p)^2-m_K^2 P^2\;.
$$
$$
\int d \rho \frac {k^{\alpha}_{\gamma}} {(p k_\gamma)}=a_{11} p^\alpha
+b_{11} P^\alpha \;.
$$
The $a_{11}$ and $b_{11}$ parameters are defined as follows
\begin{eqnarray}
a_{11}&=&-\frac 1 {(P p)^2-m_K^2 P^2}
\biggl( P^2 J_{11}- \frac {J_{12}} 2  (P p) (P^2-m_{l}^2) \biggr)\;, \nonumber
\\
b_{11}&=&\frac 1 {(P p)^2-m_K^2 P^2}
\biggl( (P p) J_{11}- \frac {J_{12}} 2  m_K^2 (P^2-m_{l}^2)\;. \biggr)\nonumber
\end{eqnarray}
\begin{eqnarray}
\int d \rho k_{\gamma}^{\alpha} &=& a_{12} P^{\alpha}\;, \nonumber 
\end{eqnarray}
where
\begin{eqnarray}
a_{12}&=&\frac {(P^2-m_{l}^2)} {2 P^2} J_{11}\;, \nonumber 
\end{eqnarray}

\begin{eqnarray}
J_1&=&\int d \rho \frac 1 {(p k_\gamma)((p_l-k_\gamma)^2-m_l^2)}=
-\frac \pi {2 I_1 (P^2-m_l^2) } \ln 
\biggl( \frac {(p p_l)+I_1} {(p p_l)-I_1} \biggr)\;, \nonumber \\
J_2&=&\int d \rho \frac 1 {(p_l-k_\gamma)^2-m_l^2}=
-\frac \pi {4 I_2} \ln \biggl( \frac {(P p_l)+I_2} {(P p_l)-I_2} \biggr)
\;, \nonumber 
\end{eqnarray}
where
\begin{eqnarray}
I_1^2&=&(p p_l)^2 -m_l^2 m_K^2\;, \nonumber\\
I_2^2&=&(P p_l)^2-m_l^2 P^2\;. \nonumber
\end{eqnarray}
\begin{eqnarray}
\int d \rho \frac {k_\gamma^\alpha} {(p_l-k_\gamma)^2-m_l^2}&=&
a_1 P^\alpha + b_1 p_l^\alpha\;, \nonumber \\
a_1&=&-\frac {m_l^2 (P^2-m_l^2) J_2+(P p_l)J_{11}} {2 ((P p_l)^2-m_l^2
P^2)} 
\;, \nonumber\\
b_1&=&\frac {(P p_l)(P^2-m_l^2) J_2+P^2 J_{11}} {2 ((P p_l)^2-m_l^2 P^2)}
\;, \nonumber
\end{eqnarray}
The integrals below are expressed in terms of the parameters, 
which can be obtained by solving
the sets of equations.
$$
\int d \rho \frac {k_\gamma^\alpha} {(p k_\gamma)((p_l-k_\gamma)^2-m_l^2)}=
a_2 P^\alpha + b_2 p^\alpha +c_2 p_l^\alpha \;,
$$
$$
\left\{
\begin{array}{ccc}
a_2 (P p)+ b_2 m_K^2+c_2 (p p_l)=J_2 \hfill \\
a_2 (P p_l)+b_2 (p p_l)+c_2 m_l^2=-\frac 1 2 J_{12} \hfill \\
a_2 P^2+b_2 (P p)+c_2 (P p_l)=(p_l q) J_1 \hfill 
\end{array}
\right.
$$
\begin{eqnarray}
\int d \rho \frac {k_\gamma^\alpha k_\gamma^\beta} 
{(p k_\gamma)((p_l-k_\gamma)^2-m_l^2)}&=&
a_3 g^{\alpha \beta}+b_3 (P^\alpha p^\beta+P^\beta p^\alpha)+
c_3 (P^\alpha p_l^\beta+P^\beta p_l^\alpha) \nonumber \\
&+& d_3 (p^\alpha p_l^\beta+p^\beta p_l^\alpha)+
e_3 p_l^\alpha p_l^\beta \nonumber \\
&+&f_3 P^\alpha P^\beta +g_3 p^\alpha p^\beta \;,\nonumber
\end{eqnarray}
$$
\left\{
\begin{array}{cccccccc}
4 a_3+2 b_3 (P p)+2 c_3 (P p_l)+2 d_3 (p p_l)+g_3 m_K^2 + e_3 m_l^2+f_3
P^2=0 \hfill \\
c_3 (p p_l) + b_3 m_K^2 + f_3 (P p)-a_1=0 \hfill \\
c_3 (P p)+d_3 m_K^2+e_3 (p p_l)-b_1=0 \hfill \\
a_3 + b_3 (P p)+d_3 (p p_l)+g_3 m_K^2=0 \hfill \\
b_3 (p p_l)+c_3 m_l^2+f_3 (P p_l)=- \frac 1 2 b_{11} \hfill \\
b_3 (P p_l)+d_3 m_l^2+g_3 (p p_l)=-\frac 1 2 a_{11}	 \hfill \\
a_3 P^2+2 b_3 P^2 (P p)+2 c_3 P^2 (P p_l)+2 d_3 (P p_l) (P p)+\hfill
\\ \hfill +e_3 (P p_l)^2+f_3 (P^2)^2+g_3 (P p)^2=(p_l q)^2J_1  \\
\end{array}
\right.
$$
$$
\int d \rho \frac {k_\gamma^\alpha k_\gamma^\beta} {(p_l-k_\gamma)^2-m_l^2}=
a_4 g_{\alpha \beta}+b_4 (P^\alpha p_l^\beta+P^\beta p_l^\alpha)+
c_4 P^\alpha P^\beta +d_4 p_l^\alpha p_l^\beta \;,
$$
$$
\left\{
\begin{array}{cccc}
a_4+d_4 m_l^2+b_4 (P p_l)=0 \hfill \\
b_4 m_l^2+c_4 (P p_l)=-\frac 1 2 a_{12} \hfill \\
4 a_4+2 b_4 (P p_l)+c_4 P^2+d_4 m_l^2=0 \hfill \\
a_4 P^2+2 b_4 P^2 (P p_l)+c_4 (P^2)^2+d_4 (P p_l)^2==\frac {(P^2-m_l^2)^2}
4 J_2 
\end{array}
\right.
$$

\newpage
\section*{Appendix 2}

\noindent
Here we present the explicit expressions for  imaginary parts of  the 
$C_i$ formfactors via the parameters, calculated in Appendix 1.
\begin{eqnarray*}
C_1&=&\frac{\alpha }{\sqrt{2}\pi} {m_{l}}(4
{a_{3}}+{b_{3}}
{m^2_{K}}+{d_{3}} {m^2_{K}}-2 {a_{2}}
{m^2_{l}}+2 {b_{3}} {m^2_{l}}-2
{c_{2}}
{m^2_{l}}+6 {c_{3}} {m^2_{l}}  \\
&&+  2 {d_{3}} {m^2_{l}}+3 {e_{3}}
{m^2_{l}}+3 {f_{3}}
{m^2_{l}}-{b_{3}}
{m^2_{\pi}}-{d_{3}} {m^2_{\pi}}-2
{b_{3}}
{(p'p_l)}  \\
&&-  2 {d_{3}} {(p'p_l)}-2 {b_{3}}
{(p'q)}-2 {d_{3}} {(p'q)}-4 {a_{2}}
{(p_lq)}+4 {b_{3}} {(p_lq)}-2 {c_{2}}
{(p_lq)}  \\
&&+ 8 {c_{3}} {(p_lq)}+2 {d_{3}}
{(p_lq)}+2 {e_{3}} {(p_lq)}+6 {f_{3}}
{(p_lq)}+2 {b_{3}} {(pp')}+2 {d_{3}}
{(pp')})\\  \\
C_5&=&-\frac{\alpha }{\sqrt{2}\pi} (4 {a_{3}}-4 {a_{2}}
{m^2_{l}}+3
{b_{3}}
{m^2_{l}}-4 {c_{2}} {m^2_{l}}+4
{c_{3}}
{m^2_{l}}+3 {d_{3}} {m^2_{l}}  \\
&&+  2 {e_{3}} {m^2_{l}}+2 {f_{3}}
{m^2_{l}}-4 {a_{2}} {(p_lq)}+4 {b_{3}}
{(p_lq)}+4 {c_{3}} {(p_lq)}+4 {f_{3}}
{(p_lq)})\\ \\
C_9&=&-\frac{\alpha }{\sqrt{2}\pi} (2 {a_{3}}+{b_{3}}
{m^2_{K}}-{a_{2}}
{m^2_{l}}+{b_{3}}
{m^2_{l}}-{c_{2}}
{m^2_{l}}+2 {c_{3}} {m^2_{l}}  \\
&&+ {d_{3}} {m^2_{l}}+2 {f_{3}}
{m^2_{l}}-{b_{3}} {m^2_{\pi}}-2 {b_{3}}
{(p'p_l)}-2 {b_{3}} {(p'q)}-2 {a_{2}}
{(p_lq)}  \\
&&+  2 {b_{3}} {(p_lq)}+2 {c_{3}}
{(p_lq)}+4 {f_{3}} {(p_lq)}+2 {b_{3}}
{(pp')})\\ \\
C_{10}&=&\frac{\alpha }{\sqrt{2}\pi}\frac{1}{{(p_lq)}}
(-{a_{1}}
{m^2_{l}}-{b_{1}} {m^2_{l}}+2
{b_{4}}
{m^2_{l}}+{c_{4}}
{m^2_{l}}+{d_{4}}
{m^2_{l}}+2 {a_{3}} {(p_lq)}  \\
&&+ {a_{2}} {m^2_{K}}
{(p_lq)}-{c_{3}} {m^2_{K}}
{(p_lq)}-{f_{3}} {m^2_{K}}
{(p_lq)}-{e_{3}} {m^2_{l}} {(p_lq)}  \\
&&+ {f_{3}} {m^2_{l}}
{(p_lq)}-{a_{2}} {m^2_{\pi}}
{(p_lq)}+{c_{3}} {m^2_{\pi}}
{(p_lq)}+{f_{3}} {m^2_{\pi}} {(p_lq)}  \\
&&- 2 {a_{2}} {(p'p_l)} {(p_lq)}+2
{c_{3}} {(p'p_l)} {(p_lq)}+2 {f_{3}}
{(p'p_l)} {(p_lq)}-2 {a_{2}} {(p'q)}
{(p_lq)}  \\
&&+ 2 {c_{3}} {(p'q)} {(p_lq)}+2
{f_{3}}
{(p'q)} {(p_lq)}+2 {f_{3}}
{{{(p_lq)}}^2}+2
{a_{2}} {(p_lq)} {(pp')}  \\
&&- 2 {c_{3}} {(p_lq)} {(pp')}-2
{f_{3}}
{(p_lq)} {(pp')}-2 {a_{2}} {(p_lq)}
{(pp_l)}+2 {b_{3}} {(p_lq)} {(pp_l)}
\\
&&+ 2 {c_{3}} {(p_lq)} {(pp_l)}+2
{f_{3}}
{(p_lq)} {(pp_l)}-2 {a_{2}} {(p_lq)}
{(pq)}+2 {b_{3}} {(p_lq)} {(pq)}  \\
&&+ 2 {c_{3}} {(p_lq)} {(pq)}+2
{f_{3}}
{(p_lq)} {(pq)})\\
\end{eqnarray*}

\begin{eqnarray*}
C_{12}&=&-\frac{\alpha }{4\sqrt{2}\pi}\frac{{m_{l}}}{
{{{(p_lq)}}^2}} (-2
{a_{12}}
{m^2_{l}}-2 {J_{11}} {m^2_{l}}-2
{a_{12}}
{(p_lq)}  \\
&&- 4 {a_{4}} {(p_lq)}+2 {J_{11}}
{(p_lq)}-{a_{11}} {m^2_{K}}
{(p_lq)}+{b_{11}} {m^2_{K}} {(p_lq)}  \\
&&+ 8 {a_{1}} {m^2_{l}} {(p_lq)}+8
{b_{1}} {m^2_{l}} {(p_lq)}-4 {b_{4}}
{m^2_{l}} {(p_lq)}-2 {c_{4}}
{m^2_{l}}
{(p_lq)}  \\
&&- 2 {d_{4}} {m^2_{l}} {(p_lq)}-4
{J_{2}} {m^2_{l}} {(p_lq)}-{b_{11}}
{m^2_{\pi}} {(p_lq)}-2 {b_{11}} {(p'p_l)}
{(p_lq)}  \\
&&- 2 {b_{11}} {(p'q)} {(p_lq)}+8
{a_{1}} {{{(p_lq)}}^2}+4 {a_{3}}
{{{(p_lq)}}^2}+4
{b_{1}} {{{(p_lq)}}^2}-4 {b_{4}}
{{{(p_lq)}}^2}\\
&&-  4 {c_{4}} {{{(p_lq)}}^2}-4 {J_{2}}
{{{(p_lq)}}^2}+2 {a_{2}} {m^2_{K}}
{{{(p_lq)}}^2}-2 {b_{2}} {m^2_{K}}
{{{(p_lq)}}^2}  \\
&&+ 2 {c_{2}} {m^2_{K}} {{{(p_lq)}}^2}+2
{g_{3}} {m^2_{K}} {{{(p_lq)}}^2}+8 {c_{3}}
{m^2_{l}} {{{(p_lq)}}^2}+6 {e_{3}}
{m^2_{l}} {{{(p_lq)}}^2}  \\
&&+ 2 {f_{3}} {m^2_{l}}
{{{(p_lq)}}^2}-2
{a_{2}} {m^2_{\pi}} {{{(p_lq)}}^2}-2 {c_{2}}
{m^2_{\pi}} {{{(p_lq)}}^2}-4 {a_{2}}
{(p'p_l)}
{{{(p_lq)}}^2}  \\
&&- 4 {c_{2}} {(p'p_l)}
{{{(p_lq)}}^2}-4
{a_{2}} {(p'q)} {{{(p_lq)}}^2}-4
{c_{2}}
{(p'q)} {{{(p_lq)}}^2}+12 {c_{3}}
{{{(p_lq)}}^3}\\
&&+ 4 {e_{3}} {{{(p_lq)}}^3}+4 {f_{3}}
{{{(p_lq)}}^3}+2 {b_{11}} {(p_lq)}
{(pp')}+4
{a_{2}} {{{(p_lq)}}^2} {(pp')}  \\
&&+ 4 {c_{2}} {{{(p_lq)}}^2} {(pp')}-2
{a_{11}} {(p_lq)} {(pp_l)}-4 {b_{11}}
{(p_lq)} {(pp_l)}+2 {J_{12}} {(p_lq)}
{(pp_l)}  \\
&&- 8 {a_{2}} {{{(p_lq)}}^2} {(pp_l)}-4
{b_{2}} {{{(p_lq)}}^2} {(pp_l)}+4
{b_{3}}
{{{(p_lq)}}^2} {(pp_l)}-8 {c_{2}}
{{{(p_lq)}}^2}
{(pp_l)}  \\
&&+ 8 {d_{3}} {{{(p_lq)}}^2} {(pp_l)}+4
{J_{1}} {{{(p_lq)}}^2} {(pp_l)}-2
{a_{11}}
{(p_lq)} {(pq)}-4 {b_{11}} {(p_lq)}
{(pq)}  \\
&&+ 2 {J_{12}} {(p_lq)} {(pq)}-4
{a_{2}}
{{{(p_lq)}}^2} {(pq)}+4 {b_{3}}
{{{(p_lq)}}^2}
{(pq)}-4 {c_{2}} {{{(p_lq)}}^2} {(pq)}
\\
&&+  4 {d_{3}} {{{(p_lq)}}^2} {(pq)})\\ \\   
C_{13}&=&-\frac{\alpha }{\sqrt{2}\pi} {m_{l}}
(2
{a_{2}}-{b_{3}}+2
{c_{2}}-2{d_{3}})\\ \\
C_{14}&=&\frac{\alpha }{\sqrt{2}\pi}\frac{
{m_{l}}}{{(p_lq)}} (2 {a_{1}}+2
{b_{1}}-4
{b_{4}}-2 {c_{4}}-2 {d_{4}}+{a_{2}}
{(p_lq)}+3 {c_{3}} {(p_lq)}  \\
&&+ 2 {e_{3}} {(p_lq)}+{f_{3}}
{(p_lq)})
\end{eqnarray*}
\begin{eqnarray*}
C_{16}&=&\frac{\alpha }{4\sqrt{2}\pi}\frac{1}{ {{{(p_lq)}}^2}}(-4
{a_{12}}
{m^2_{l}}-4
{J_{11}} {m^2_{l}}-4 {a_{12}} {(p_lq)}-8
{a_{4}} {(p_lq)}+4 {J_{11}} {(p_lq)}
\\
&&-  2 {a_{11}} {m^2_{K}} {(p_lq)}+16
{a_{1}} {m^2_{l}} {(p_lq)}+16 {b_{1}}
{m^2_{l}} {(p_lq)}+{b_{11}}
{m^2_{l}}
{(p_lq)}  \\
&& - 8 {b_{4}} {m^2_{l}} {(p_lq)}-4
{c_{4}} {m^2_{l}} {(p_lq)}-4 {d_{4}}
{m^2_{l}} {(p_lq)}-8 {J_{2}}
{m^2_{l}}
{(p_lq)}  \\
&&+  16 {a_{1}} {{{(p_lq)}}^2}+4 {a_{3}}
{{{(p_lq)}}^2}+8 {b_{1}} {{{(p_lq)}}^2}-8
{b_{4}}
{{{(p_lq)}}^2}-8 {c_{4}} {{{(p_lq)}}^2}  \\
&&-  2 {J_{12}} {{{(p_lq)}}^2}-8 {J_{2}}
{{{(p_lq)}}^2}-4 {b_{2}} {m^2_{K}}
{{{(p_lq)}}^2}+4 {g_{3}} {m^2_{K}}
{{{(p_lq)}}^2}  \\
&& - 2 {a_{2}} {m^2_{l}}
{{{(p_lq)}}^2}-2
{c_{2}} {m^2_{l}} {{{(p_lq)}}^2}+4 {c_{3}}
{m^2_{l}} {{{(p_lq)}}^2}+4 {e_{3}}
{m^2_{l}} {{{(p_lq)}}^2}  \\
&&- 4 {a_{2}} {{{(p_lq)}}^3}+4 {c_{3}}
{{{(p_lq)}}^3}-2 {a_{11}} {(p_lq)}
{(pp_l)}-4
{b_{11}} {(p_lq)} {(pp_l)}  \\
&&+ 4 {J_{12}} {(p_lq)} {(pp_l)}-8
{a_{2}} {{{(p_lq)}}^2} {(pp_l)}-8
{b_{2}}
{{{(p_lq)}}^2} {(pp_l)}+4 {b_{3}}
{{{(p_lq)}}^2}
{(pp_l)}  \\
&&-  8 {c_{2}} {{{(p_lq)}}^2} {(pp_l)}+8
{d_{3}} {{{(p_lq)}}^2} {(pp_l)}+8
{J_{1}}
{{{(p_lq)}}^2} {(pp_l)}-2 {a_{11}}
{(p_lq)}
{(pq)}  \\
&&- 4 {b_{11}} {(p_lq)} {(pq)}+4
{J_{12}} {(p_lq)} {(pq)}+4 {b_{3}}
{{{(p_lq)}}^2} {(pq)})
\end{eqnarray*}

\vspace*{2cm}
\begin{displaymath}
C_3=C_7=C_{11}=C_{15}=0
\end{displaymath}

\newpage

\section*{Figure captions}

\vspace*{1.5cm}
\noindent
\begin{description}
\item[Fig. 1.] The Feynman diagrams for the  $K^{+}\to \pi^0 l^{+}\nu\gamma$ decay 
at the tree level of SM.
\item[Fig. 2.] Branching differential distribution over the 
pion, lepton, $\gamma$-quantum momenta and the angle between 
the lepton and $\gamma$-quantum in the $K$-meson rest frame for the
cases of (a) electron and (b) muon decay channels. 
\item[Fig. 3.] $\xi$-dependence of the  $K^{+}\to \pi^0 l^{+}\nu\gamma$ branching
at the tree level of SM for the (a) electron and (¡) muon channels.
\item[Fig. 4.] The Feynman diagrams contributing
to the imaginary parts of formfactors (10) at the one-loop level of 
SM.
\item[Fig. 5.] $\xi$-odd contributions, $f_{odd}$, to the 
branching differential distribution¢  for the  
(a) electron and (b) muon decay channels.
\end{description}

\newpage
\setlength{\unitlength}{1mm}
\begin{figure}[ph]
\begin{picture}(150, 200)
\put(-10,100){\epsfxsize=12cm \epsfbox{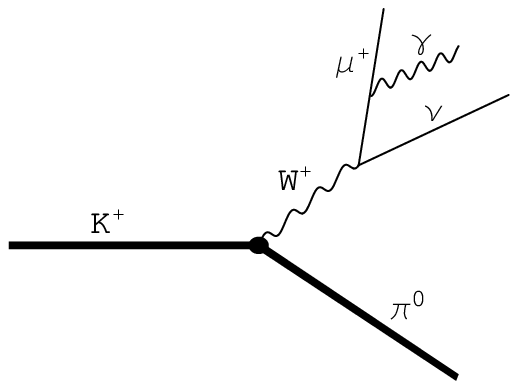}}
\put(80,100){\epsfxsize=12cm \epsfbox{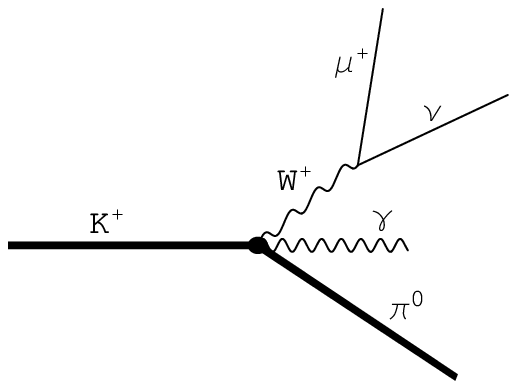}}
\put(75,90){\large\bf{Fig. 1}}
\end{picture}
\end{figure}

\newpage
\setlength{\unitlength}{1mm}
\begin{figure}[ph]
\bf
\begin{picture}(150, 200)

\put(-40,20){\epsfxsize=16cm \epsfbox{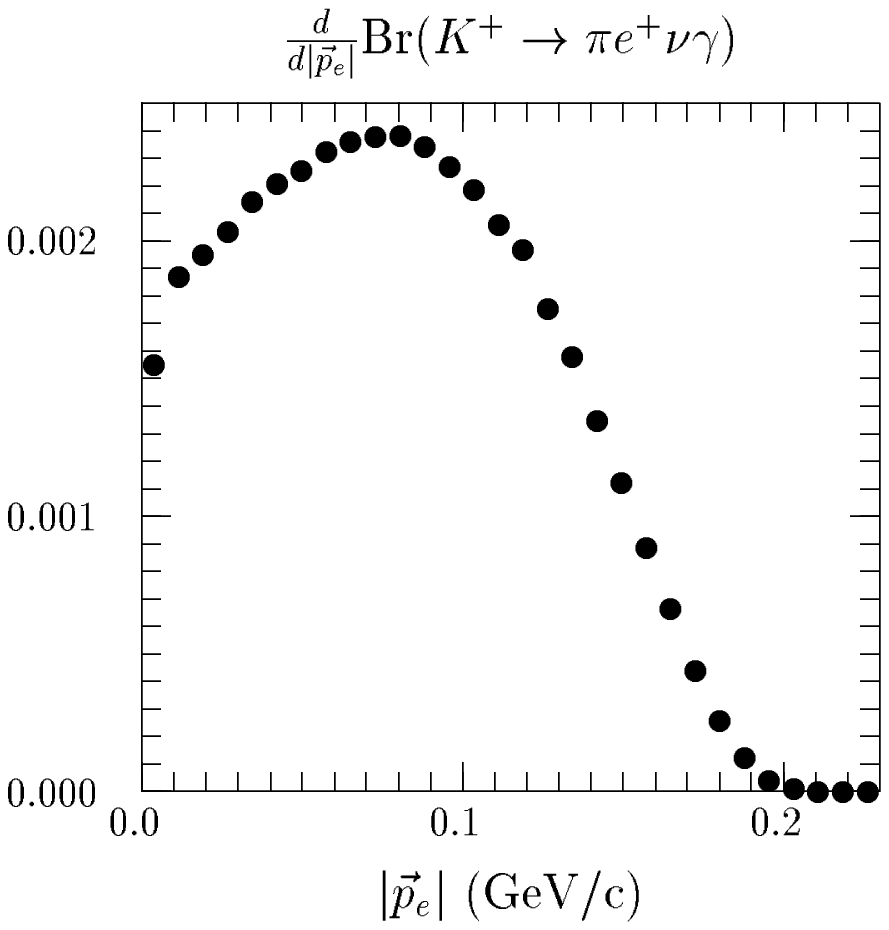}}

\put(45,20){\epsfxsize=16cm \epsfbox{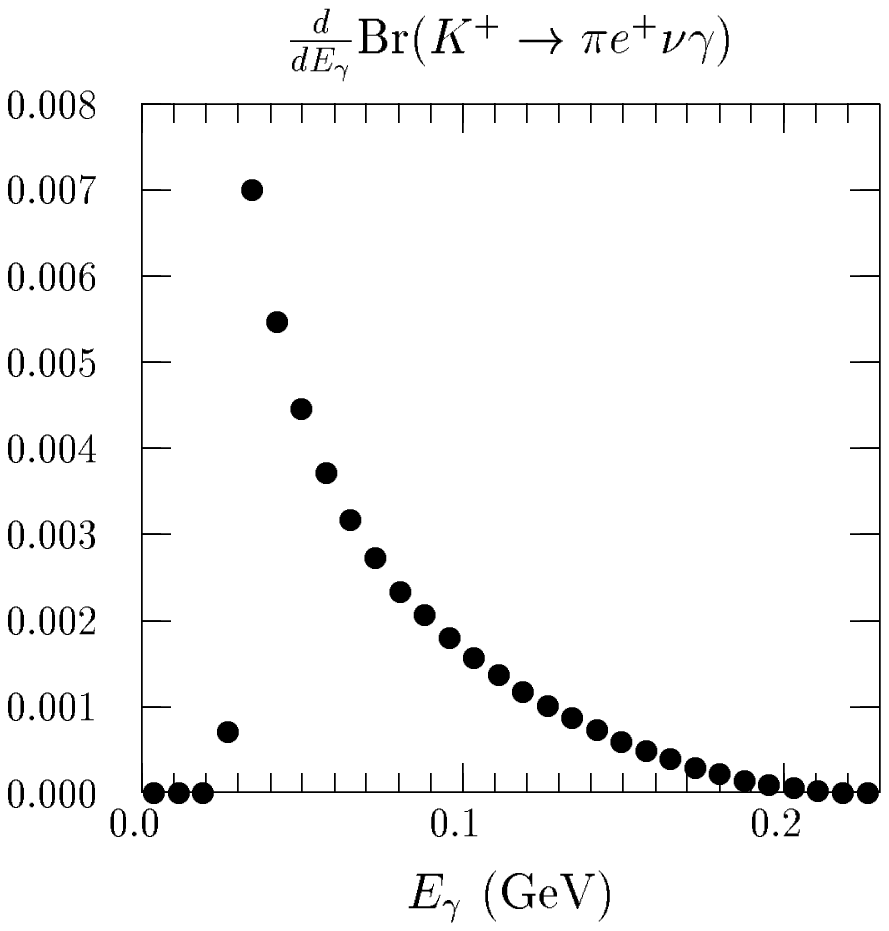}}

\put(-40,-70){\epsfxsize=16cm \epsfbox{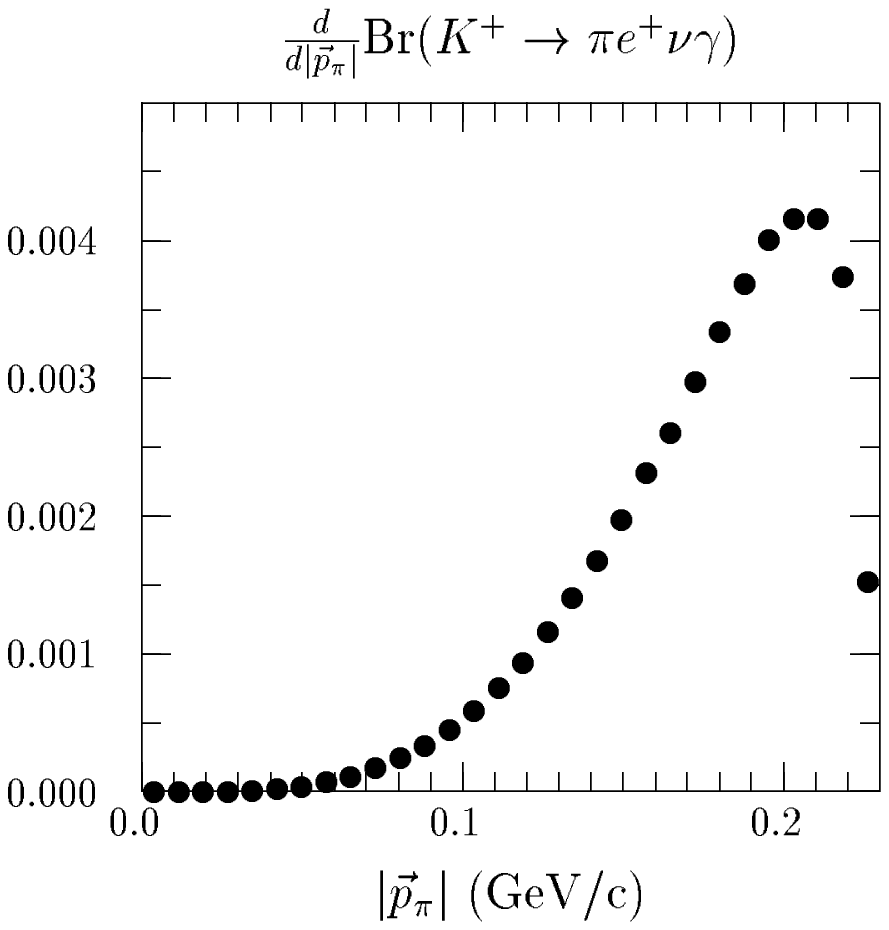}}

\put(45,-70){\epsfxsize=16cm \epsfbox{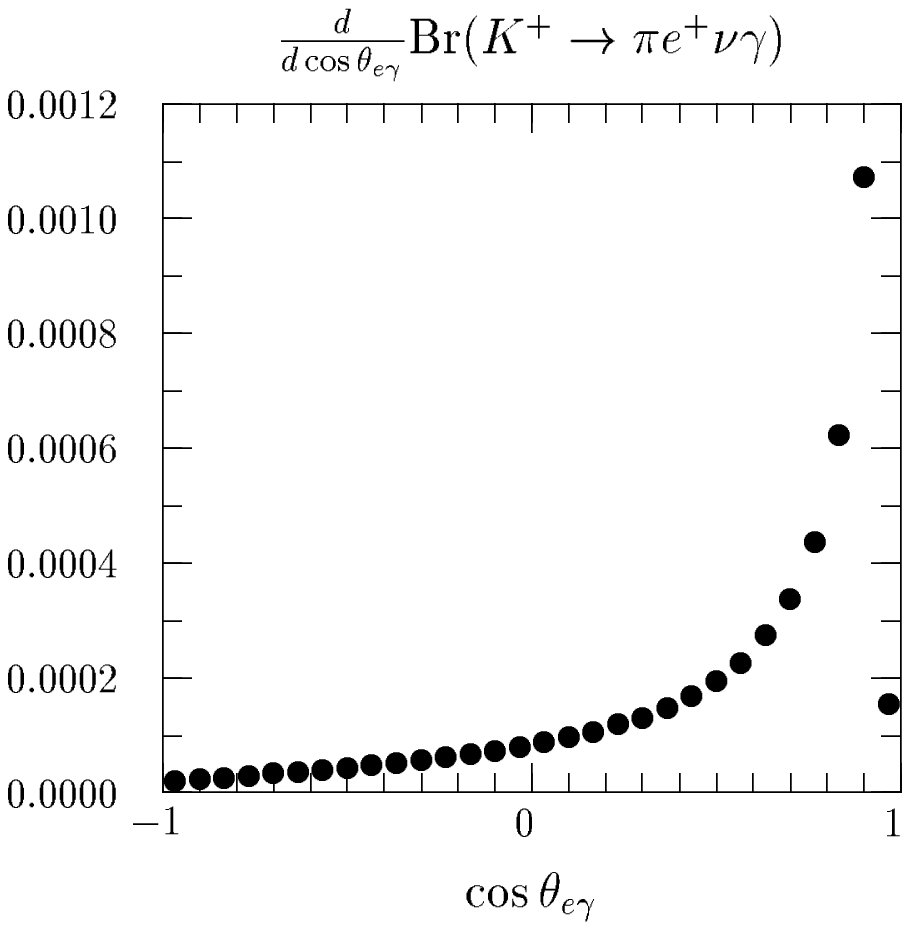}}
\put(70,10){Fig. 2a}
\end{picture}
\end{figure}

\newpage
\setlength{\unitlength}{1mm}
\begin{figure}[ph]
\bf
\begin{picture}(150, 200)

\put(-40,20){\epsfxsize=16cm \epsfbox{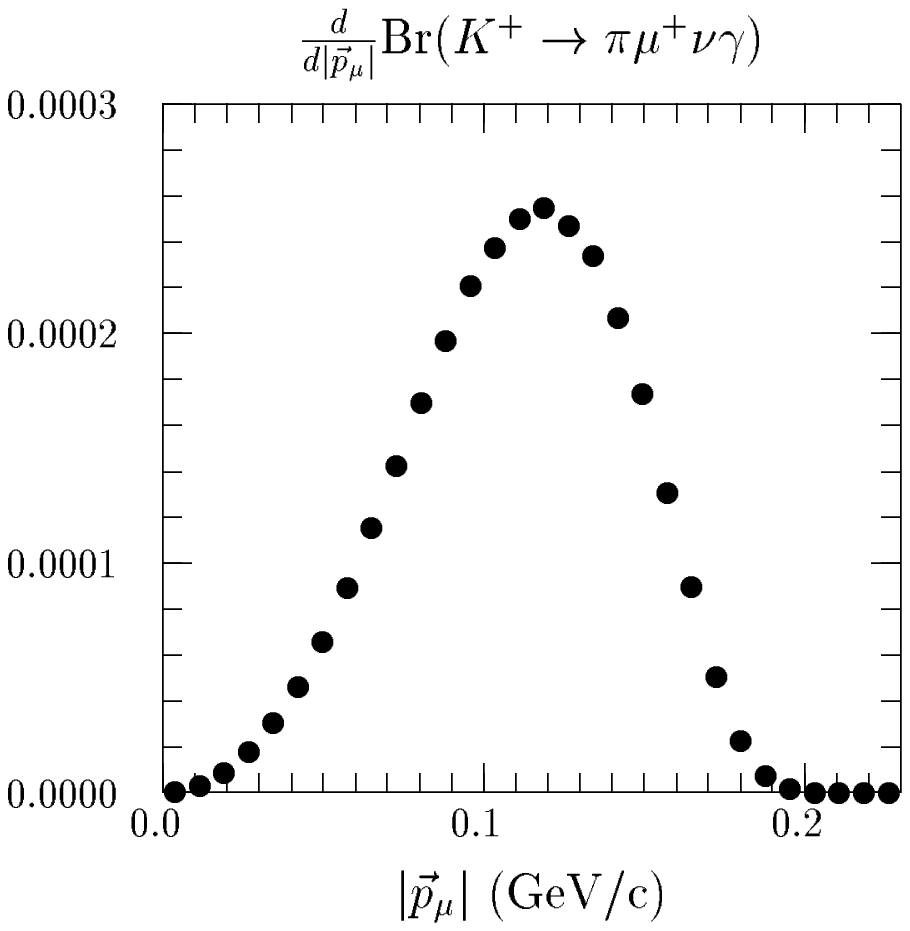}}

\put(45,20){\epsfxsize=16cm \epsfbox{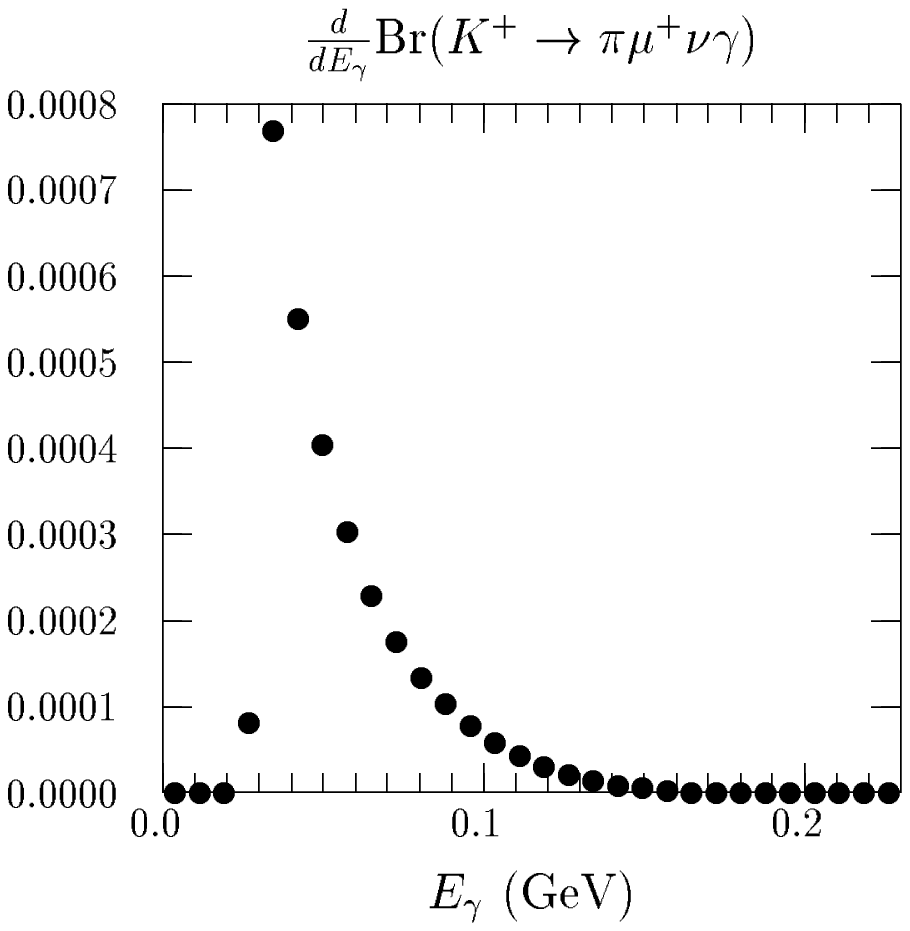}}

\put(-40,-70){\epsfxsize=16cm \epsfbox{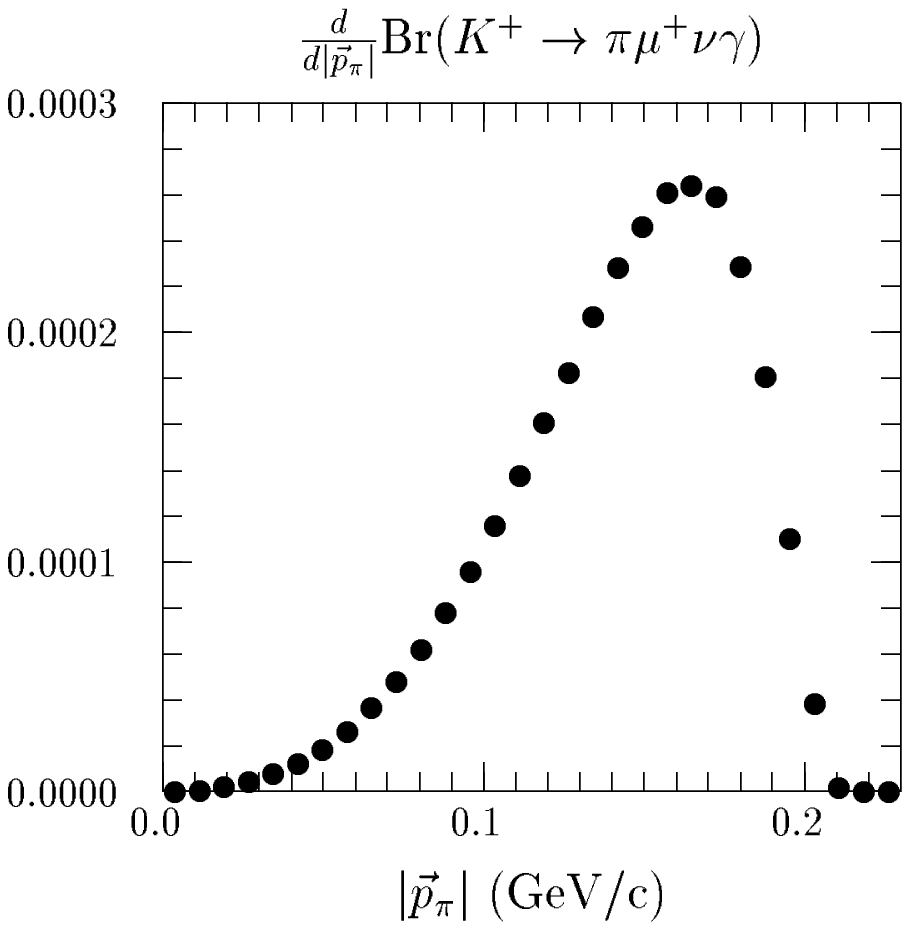}}

\put(45,-70){\epsfxsize=16cm \epsfbox{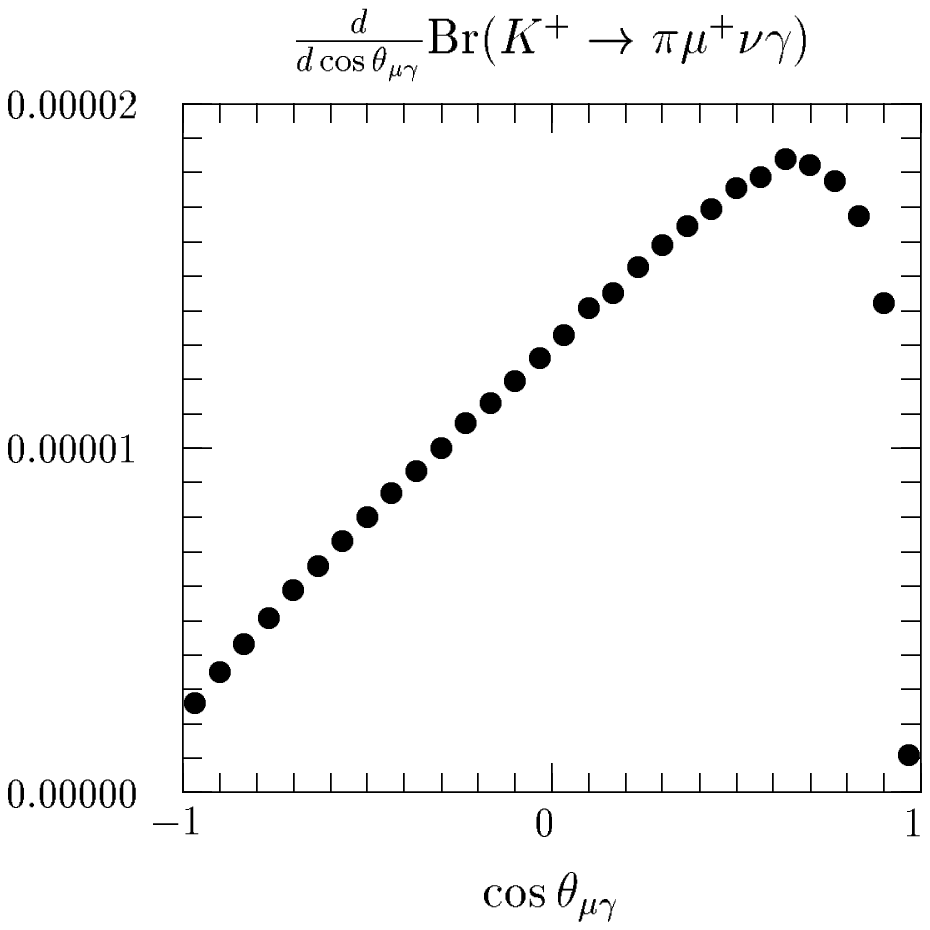}}
\put(70,10){Fig. 2b}
\end{picture}
\end{figure}

\newpage
\setlength{\unitlength}{1mm}
\begin{figure}[ph]
\bf
\begin{picture}(150, 200)

\put(0,0){\epsfxsize=18cm \epsfbox{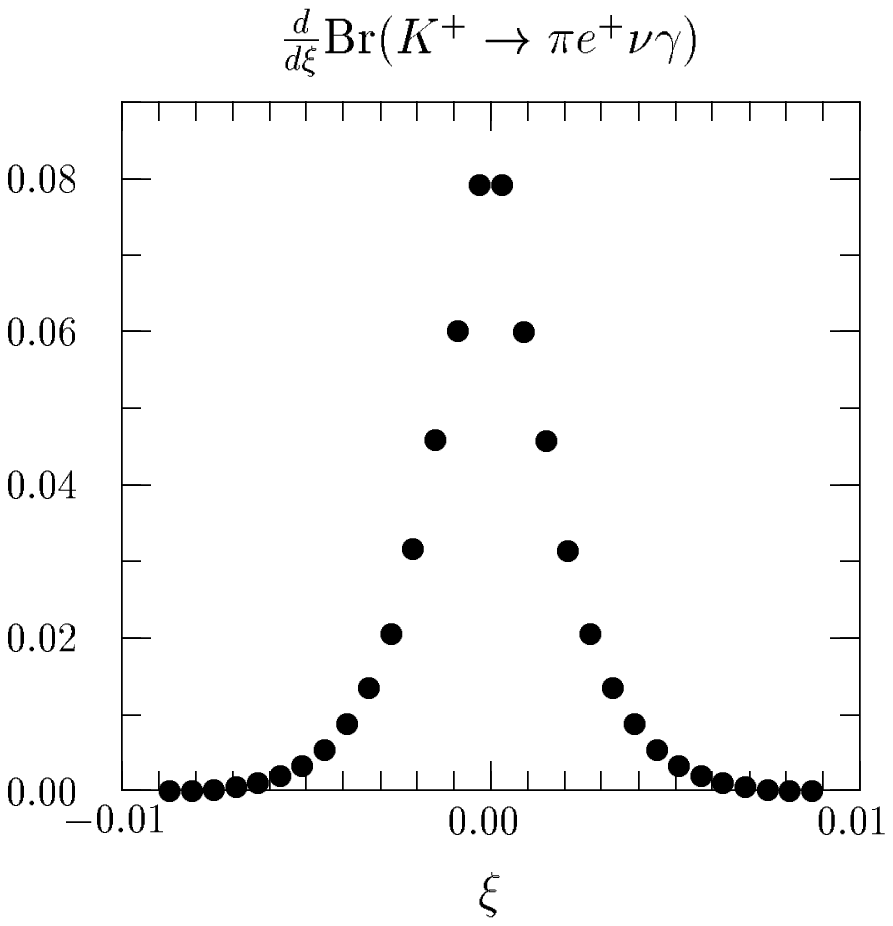}}
\put(80,110){Fig. 3a}

\put(0,-100){\epsfxsize=18cm \epsfbox{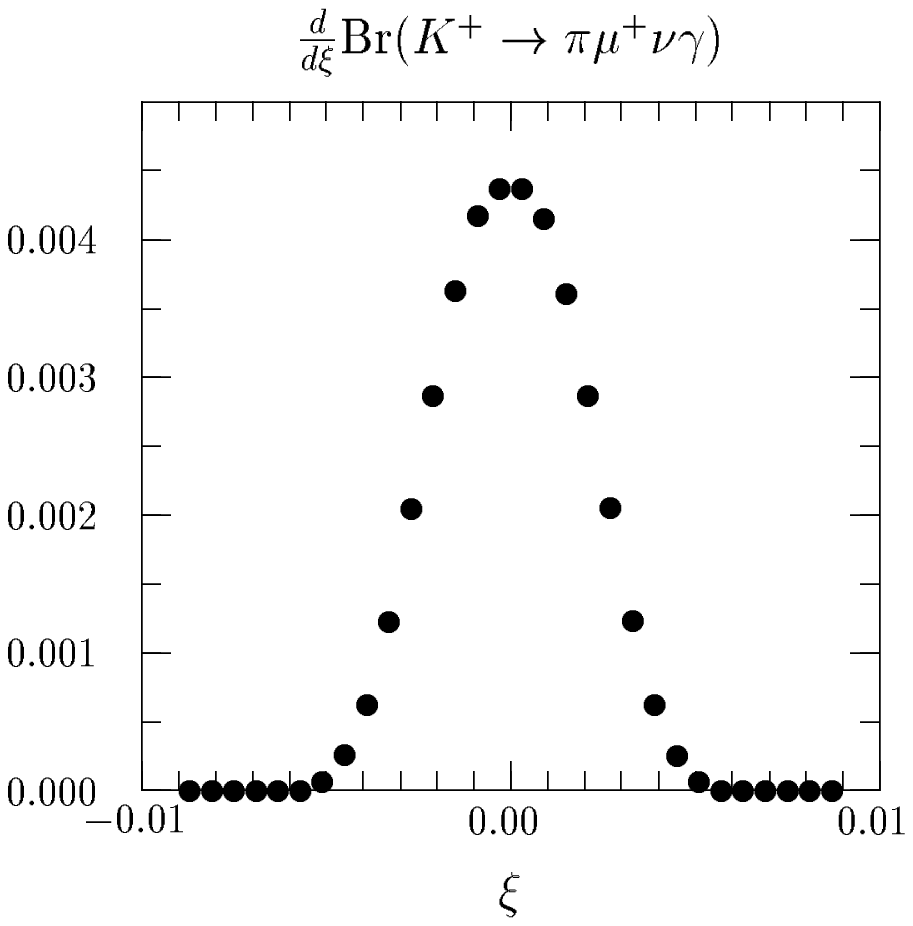}}

\put(80,10){Fig. 3b}
\end{picture}
\end{figure}

\newpage
\setlength{\unitlength}{1mm}
\begin{figure}[ph]
\begin{picture}(150, 200)
\put(-10,150){\epsfxsize=12cm \epsfbox{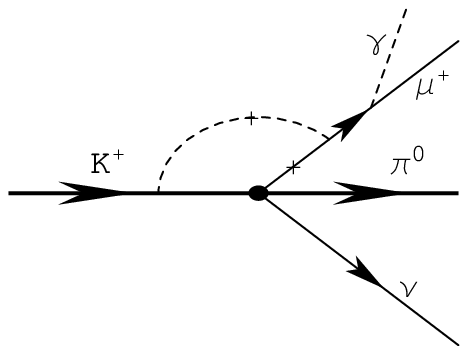}}
\put(80,150){\epsfxsize=12cm \epsfbox{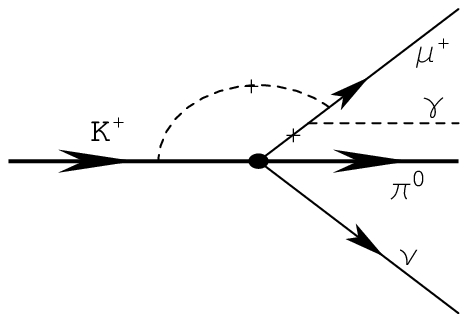}}
\put(-10,70){\epsfxsize=12cm \epsfbox{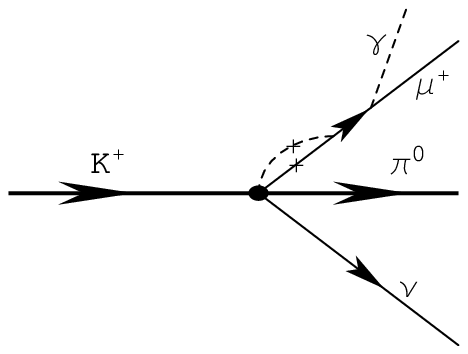}}
\put(80,70){\epsfxsize=12cm \epsfbox{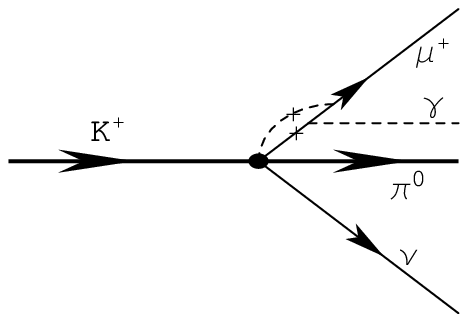}}
\put(-10,-0){\epsfxsize=12cm \epsfbox{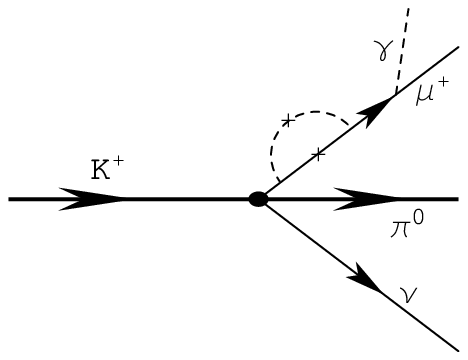}}
\put(80,-0){\epsfxsize=12cm \epsfbox{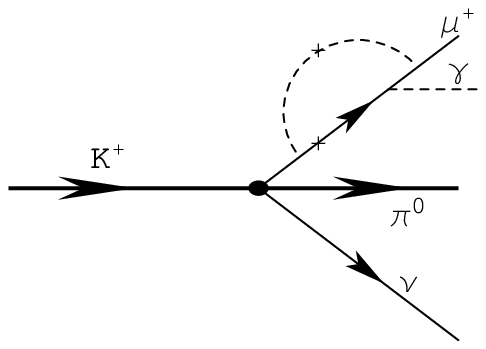}}

\put(25,170){\large\bf (a)}
\put(25,90){\large\bf (c)}
\put(25,15){\large\bf (e)}

\put(100,170){\large\bf (b)}
\put(100,90){\large\bf (d)}
\put(100,15){\large\bf (f)}

\put(70,0){\large\bf{Fig. 4}}
\end{picture}
\end{figure}

\newpage
\setlength{\unitlength}{1mm}
\begin{figure}[ph]
\begin{picture}(150, 200)
\put(20,130){\epsfxsize=10cm \epsfbox{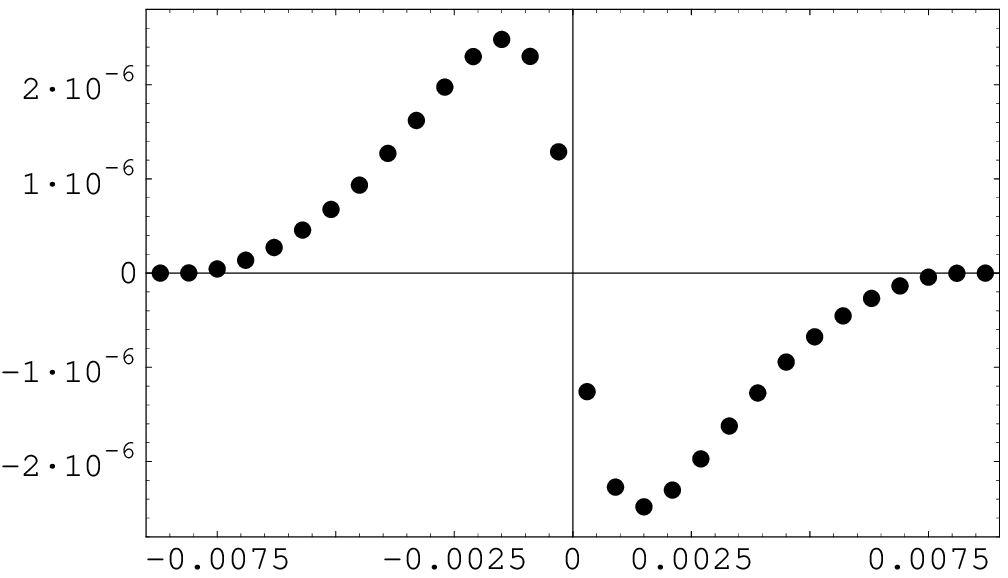}}
\put(20,30){\epsfxsize=10cm \epsfbox{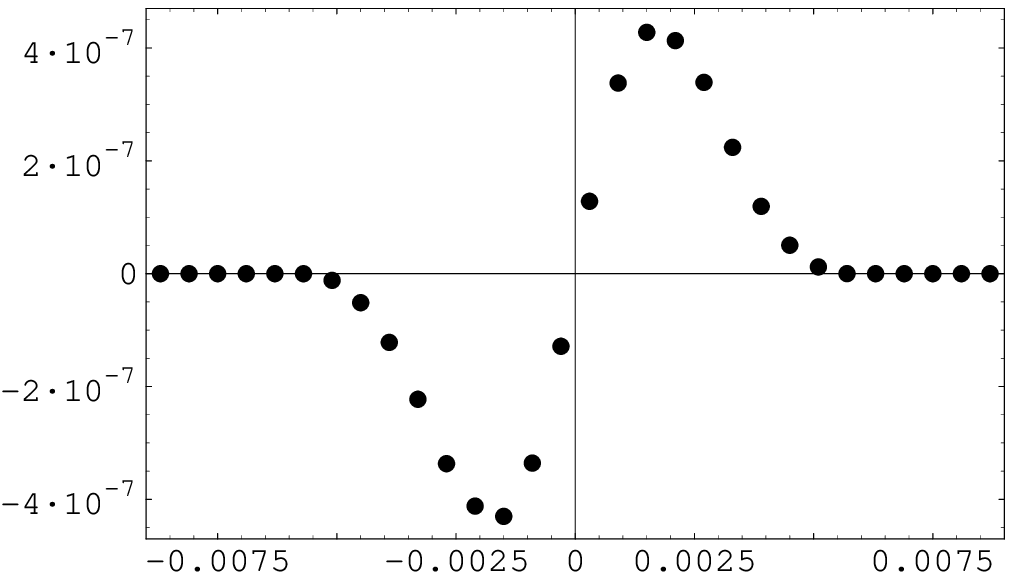}}
\put(55,95){$\frac{d}{d\xi}\hbox{Br}(K^+\to\pi \mu^+ \nu \gamma)$}
\put(75,27){$\xi$}
\put(55,195){$\frac{d}{d\xi}\hbox{Br}(K^+\to\pi e^+ \nu \gamma)$}
\put(75,127){$\xi$}
\put(70,110){\large\bf{Fig. 5a}}
\put(70,10){\large\bf{Fig. 5b}}
\end{picture}
\end{figure}

\end{document}